\documentclass[onecolumn]{article}        \usepackage{graphicx}
\usepackage{adjustbox}
\usepackage{amsmath}
\usepackage{epstopdf}
\usepackage[graphicx]{realboxes}
\usepackage{rotating}
\usepackage{float}
\usepackage{latexsym}
\usepackage{amssymb}
\usepackage{array}
\usepackage{longtable}
\usepackage{dcolumn}
\usepackage{bm}
\usepackage{subcaption}
\usepackage{caption}
\begin{document}

\title{Cosmological models with Bounce scenario in f(Q,C)-gravity}

\author{Bhaswati Mandal and B. C. Paul \\
Physics Department, North Bengal University \\
 Dist : Darjeeling, Pin : 734013, India\\
 Email : bcpaul@nbu.ac.in, bhaswatimandal93@gmail.com}

\date{Received: date / Accepted: date}

\maketitle

\vspace{0.5in}

\begin{abstract}

We present a bounce universe in modified $f(Q,C)$ gravity considering linear as well as exponential form of gravity. Bounce cosmological models are introduced to remove the singularity problem of the early universe. A new quadratic boundary term ($C^2$), which is added in the modified gravity to study different features of the universe in the framework of bouncing cosmology. Both power law expansion and exponential emergent universe are explored in linear  modified gravity. The energy conditions and stability of cosmological bounce are investigated. We also compared power law expansion in linear and exponential form of modified gravity.  
\end{abstract}

\section{ Introduction}
\label{intro}

In modern cosmology it is accepted that the present universe emerged from an inflationary phase in the past \cite{1}.
Inflation solves several cosmological issues that encounters in the Standard Big Bang cosmology, and benefit out of it is that inflationary scenario describes a causal theory of structure formation \cite{2}.
In spite of the phenomenological success, inflationary theory encountered a number of conceptual challenges \cite{3}  which motivated the exploration of an alternative  scenario of the early universe as the evolution before inflation may decelerating in nature and once again singularity problem cropped up. 
Recent astronomical  observations  namely, the type Ia supernovae DATA \cite{9,10,11}, Cosmic Microwave Background Radiation \cite{12} and large-scale structure observations \cite{13,14,15} predicted that the present universe is not only expanding it is accelerating.
 Although the  general relativity (GR) is the most successful fundamental gravitational theory which describes the large-scale structure of the universe  fairly well, it fails to accommodate the late universe unless the gravitational and/or matter sector(s) of the Einstein field equations (EFE) are modified. In modern cosmology it is believed  that the universe is dynamically evolving through two different phases of accelerated expansion, at the beginning of the universe, it emerged from an inflationary epoch and the second phase of acceleration can be realized very recently from supernova type-Ia prediction. The observational analysis predict  that the universe is not only expanding it is accelerating, which is called late acceleration. The later feature of evolution fits  well with observation in a Lambda cold dark matter or $\Lambda$CDM model. But the $\Lambda$CDM model is found to have some conceptual issues like fine-tuning and cosmic coincidence problems \cite{16, 17, 18, 19}. Therefore, it is necessary to look for alternative approaches in order to generalize the fundamental theories of gravity accommodating both the expansion.

Since GR with normal matter cannot accommodate the present acceleration of the universe, an alternative is the modification of the matter sector of the EFE with different dynamical Dark Energy (DE), namely, models consisting of one or more scalar fields namely, quintessence \cite{20, 21, 22}, Chaplygin gas \cite{23} and its variations \cite{24, 25}. On the other hand modifications in the gravitational sector, namely $f(R)$ theories of gravity \cite{26, 27} where $R$ is the Ricci scalar, $f(R, T)$ gravity \cite{28} with $T$ being the trace of the energy-momentum tensor, modified Gauss-Bonnet gravity \cite{29, 30, 31}, Brane world gravity \cite{32, 33}, Horava-Lifshitz theory of gravity \cite{34}. 

Recently, another proposal to  modify the gravity making use of the torsion surfaced which is called the teleparallel equivalent of general relativity (TEGR) \cite{44}. In TEGR, the curvature and non-metricity are zero and in this case the Weitzenbck connection is the affine connection \cite{46, 47, 48}. The fundamental objects are tetrads by which one can derive the affine connection, the torsion invariants, and that led to the modification of field equations. It is found that TEGR is modified similar to the modification of Einstein-Hilbert (EH) action that finally based on curvature formulation (Leading to an action that yields the Einstein field equations). The simplest modification is the $f(\mathcal{T})$ gravity \cite{49}, where $\mathcal{T}$ is the torsion scalar.  In $f(\mathcal{T})$ gravity, the field equations are of second order differential equation, unlike in $f(R)$ gravity, which in the metric approach is a fourth-order theory. $f(\mathcal{T})$ gravity theories have been applied to explore astrophysical objects and cosmic observations, and in particular, they are extensively used to explain the late-time acceleration without dark energy. \cite{49, 50, 51, 52, 53,54} (For details on teleparallel theories Ref.\cite{57})). 

It is also known that equivalent formulation of GR can be constructed making use of non-metricity in which a flat spacetime is considered with total curvature and vanishing torsion, known as symmetric teleparallel equivalent of GR (STEGR) \cite{58, 59, 60, 61}. Similar to $f(R)$ theories, an extension of STEGR is $f(Q)$ gravity \cite{62}, where modifications on a Lagrangian density is incorporated making use of a  non-metricity scalar $Q$. However when $f(Q) = Q$ one recovers STEGR. The class of theories obtained from the generalized Friedmann equation is found to describe the late accelerated expansion of the universe. The modified $f(Q)$ gravity is studied phenomenologically in the context of cosmology \cite{63, 64, 65, 66, 67, 68}. It is shown that the $f(Q)$ theory is one of the most promising alternative theory of gravity for interpreting cosmological observations \cite{69, 70, 71, 72, 73}. In the literature \cite{64} scalar, vector, and tensor modes of perturbations have been studied in the framework of $f(Q)$ gravity, which is supported by the constraints imposed by Big Bang Nucleosynthesis (BBN) \cite{74}.

Although it is known that the present universe is emerged from an inflationary epoch in the past, it is pertinent to investigate the evolutionary phase of the universe before the inflation. In this scenario before the early era the kinetic energy dominated era definitely encounter a singularity epoch. To avoid the singularity that one encounters in the framework of scalar field cosmology  cosmologists look for a bouncing scenario or a singularity free universe in nonflat \cite{BCP} and flat universe \cite{bcp,bcp1,bcp2}. It is known that  a regular bounce with flat metric can be obtained in GR provided the NEC ($\rho+p \geq 0$) is violated. It is also shown that such bounce solution can be obtained with negative energy scalar field; it is shown that the bounce solution with negative energy scalar fields is permissible \cite{4}, ghost condensates \cite{5}, conformal Galileon field \cite{6}. However, it is found that the bounce obtained above led to instabilities \cite{7}. Recent times bouncing cosmologies are explored in the framework of $f(R,T)$ gravity to avoid the big bang singularity \cite{75}, where $T$ is the trace of the matter tensor.

There are other proposal in cosmology where  boundary term ($C$) plays an important role \cite{8} in symmetric telleparallel gravity with a smooth function of modified gravity $f(Q)$ using the Gibbons-Hawking-York (GHY) boundary terms \cite{9}. The  conservation equation is obtained and the validity of GR field equation is found to satisfy on a manifold with boundary term.
The $f(Q,C)$-theory is an extension of  symmetric teleparallel gravity endowed with some interesting properties. The metric affine theory $f(Q)$ is a second order theory which is elevated to fourth order in the field equation of $f(Q,C)$-gravity. Furthermore the theory is consistent with $f(Q)$-gravity in the limit $C\rightarrow 0$. The motivation of the paper is to look for  the role of the boundary term in obtaining cosmological model for accommodating bounce solutions at the early epoch. Thus, the modified theory $f(Q,C)$-gravity is introduced in recent times to investigate the nature of DE. The inclusion of the boundary term $C$ in the modified gravity is a novel idea to accommodate the observed late universe.

The paper is presented as follows : in Sec 1, we present introduction and motivation of the paper, in Sec. 2.Gravitational action and the field equations are derived, in Sec 3. we present different bouncing cosmological models. In sec. 4 stability of de Sitter like solution and conclusion in Sec. 5.Discussion.

\section{Gravitational Action and  Field Equations}

The gravitational action of the $f(Q,C)$ -gravity  is given by
\begin{equation}
\label{e1}
I=  \int \left( \frac{1}{16 \pi G}   \; f(Q,C)  + \mathcal{L}_m  \right)  \sqrt{-g}\; d^4x \end{equation}
where $C=R-Q$, and $R$ is the Ricci scalar $G$ is gravitational  constant and $\mathcal{L}_m$ represents the matter Lagrangian.
The non-metricity term is
\begin{equation}
\label{e2}
Q=-g^{\mu \nu} \left( L^{\alpha}_{\sigma \mu} L^{\sigma}_{ \nu \alpha}-L^{\alpha}_{\sigma \alpha} L^{\sigma}_{\mu \nu}  \right)
\end{equation}
where the deformation tensor is
\begin{equation}
\label{e3}
L^{\phi}_{\mu  \nu} = -\frac{1}{2} g^{\phi \xi}    \left( \nabla_{\nu} g_{ \mu \alpha}-\nabla_{\mu} g_{ \xi \phi} + \nabla_{\phi} g_{ \mu\nu} \right)
\end{equation}
The superpotential is
\begin{equation}
\label{e4}
P^{\alpha}_{\mu  \nu} = -\frac{1}{2} L^{\alpha}_{\mu \nu }   + \frac{1}{4} ( Q^{\alpha} - \tilde{Q}^{\alpha} )  g_{ \mu \nu} -
\frac{1}{4}  \delta^{\alpha}   (_{\mu} Q_ {\nu })
\end{equation}
The non-metricity scalar is determined by the superpotential which becomes
\begin{equation}
\label{e5}
Q= - Q_{\alpha \beta \gamma} P^{\alpha \beta \gamma} =  - \frac{1}{4} \left[ -Q^{\alpha \beta \gamma} Q_{\alpha \beta \gamma} 
+2 Q^{\alpha \beta \gamma} Q_{\gamma\alpha \beta} -2 Q^{\xi}  \tilde{Q}_{\xi} + Q^{\xi}Q_{\xi} \right]
\end{equation}
The field equation can be derived by the variation of the action (1) which yields
\[
\kappa T_{\mu \nu}   = - \frac{f}{2}    g_{\mu \nu} + \frac{2}{\sqrt{-g} }     \partial_{\alpha} \left( \sqrt{-g} f_Q P^{\alpha}_{\mu \nu} \right)
  + \left(  P_{\mu \gamma \beta} Q^{\gamma \beta} _{\nu}    - 2 P_{\gamma \beta \nu}  Q^{\gamma \beta}_{mu}  \right) f Q 
  \]
\begin{equation}
\label{e6}  
\;\;\;\;\;\;\;\;\;
   + 
  \left( \frac{C}{2} g_{\mu \nu} - \nabla_{\mu}  \nabla_{\nu} +g_{\mu \nu} \nabla^{\gamma}  \nabla_{\gamma}-2 P^{\alpha}_{\mu \nu}  \partial_{\alpha}   \right) f_C
\end{equation}

 We consider a flat, homogeneous and isotropic universe  described by Friedmann-Robertson-Walker metric given by
\begin{equation}
\label{e7}
ds^{2}= - dt^{2} + a(t)^2 \left( dr^{2}+ r^2 (d\theta^2 + sin^2 \theta d\phi^2) \right).
\end{equation}

The isotropic matter  configuration is given by
\begin{equation}
\label{e8}
T_{\mu \nu} = (\rho+p) u_{\mu } u_{\nu } -p g_{\mu \nu}
\end{equation}
where $\rho$, $p$ and $u_{\mu}$ are  the energy density, pressure and four velocity of the fluids respectively.
We define Hubble parameter $H= \frac{\dot{a}}{a}$, where over dot represents cosmic time derivative, $R$ the Ricci scalar, $Q$ is the non-metricity term and the Boundary term $C$ are given by
\[
R= 6 (2H^2 + \dot{H}),
\]
\[
Q=-6H^2,
\]
\begin{equation}
\label{e9}
C=R-Q= 6 (\dot{H} +3 H^2).
\end{equation}
Using eqs. (\ref{e6}) and  (\ref{e8})  we obtain the resulting field equations
\begin{equation}
\label{e10}
\rho= \frac{1}{2} f +6H^2 f_Q - (9H^2+3 \dot{H} ) f_C + 3H \dot{f}_{C}
\end{equation}
\begin{equation}
\label{e11}
p= -\frac{1}{2} f - (6H^2+2 \dot{H} ) f_Q -2H\dot{f}_Q + (9H^2 +3 \dot{H}) f_C - \ddot{f}_C
\end{equation}
above we put $8 \pi G =1$. The Null energy condition(NEC), Dominant energy condition(DEC), Strong energy condition (SEC) are expressed as
 \begin{equation}
\label{e12}
\rho +p =   - 2H\dot{f}_Q    - 2 \dot{H}  f_Q  + 3H \dot{f}_{C}  - \ddot{f}_C
\end{equation}
 \begin{equation}
\label{e13}
\rho -p =  f+ 6 H^2 f_Q+ - 2H \dot{f}_Q - 2(9 H^2+ 3 \dot{H}) f_C+ 3H \dot{f}_{C}  +(6 H^2+ 2\dot {H}) f_Q + \ddot{f}_C
\end{equation}
\begin{equation}
\label{e14}
\rho +3 p =  -f+ 6 H^2 f_Q - 6 H \dot{f}_Q + 2(9 H^2+ 3 \dot{H}) f_C+ 3H \dot{f}_{C} - 3 \ddot{f}_C
\end{equation}
  The equation being pressure of the fluid can be determined for given form of $f(Q,C)$ gravity, which will be studied to obtain bouncing state in the next section.
  
\section {Bouncing Universe Model } 
A bouncing  model of the universe can be obtained when the model parameters satisfy the  following  criterion:\\
$ \bullet $ In GR bouncing  solutions violate the null energy condition  (NEC)  in the vicinity of the bouncing point, which permits when the Hubble parameter $H= - 4 \pi G \rho (1+ \omega) >0$. \\
$ \bullet $ During  the contraction phase of the universe,  the scale factor $a(t)$ decreases $ i.e. $  $a(t)<0$  and the corresponding Hubble parameter, $H(t)<0$, but during the expansion phase of the universe, the scale factor must grow {\it i.e.} $a(t)>0$  and $H(t)>0$.  However,  at the bouncing point  $a(t)=0$ and $H(t)=0$ at $t=0$.\\
$\bullet $ The  equation  of   state  (EoS)  parameter $\omega$ crosses the quintom line (phantom divide) $\omega = -1$  in the vicinity of the bouncing point.

\subsection{Model  I:  Power Law Model} 
  We consider a scale factor of the universe given by
 \begin{equation}
\label{e15}
a(t)= a_o \left( 1+ \frac{3 \sigma }{2} \; t^2 \right)^{\frac{1}{3}}.
\end{equation}
 where $a_0$ and $\sigma$ are arbitrary constants and the scale factor satisfies the criterion for a bouncing universe. The Hubble parameter  for the  scale factor is given by
 \begin{equation}
\label{e16}
H(t) = \frac{2 \sigma \; t}{2 + 3\sigma \; t^2}.
\end{equation}
The scale factor eq. (\ref{e15}),  ensures  a bouncing scenario at $t=t_c=0$, it satisfies the bouncing condition $H<0$ for $t<t_c$, $H>0$ for $t> t_c$ and $H=0$ at $t=t_c$. At the bouncing point one obtains $\dot{H}=\frac{1}{2} \sigma >0$. 
The deceleration parameter for bouncing cosmological  solution is  given by
 \begin{equation}
\label{e17}
q(t)= -1+ \frac{d}{dt} \left(H^{-1}\right) = \frac{1}{2} - \frac{1}{\sigma \; t^2}. 
\end{equation}
A consistent trend of behavior of the evolution of the deceleration parameter is found in this case for different $\sigma$, it satisfies a  limit $-\frac{\sqrt{2}}{\sqrt{\sigma}}$ $<t_c < $ $\frac{\sqrt{2}}{\sqrt{\sigma}}$ for obtaining the bouncing epoch. 

\begin{figure}[H]
\centering
\includegraphics[width=8.41cm,height= 5.5 cm]{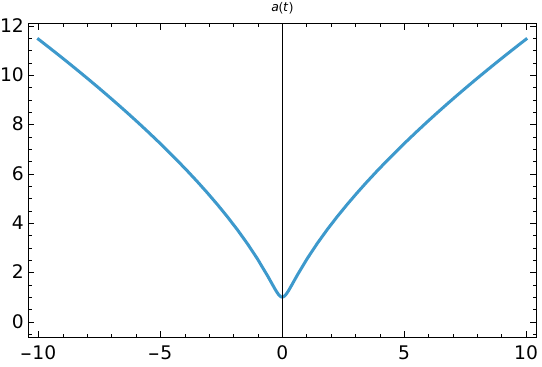}
\caption{Evolution of Hubble parameter versus cosmic time  for $\sigma=10 $ }
\end{figure}

\vspace{0.5 cm}

For a given $\sigma$, the time evolution of the scale factor and the corresponding Hubble parameter are drawn in Figs. (1) and (2) respectively. The scale factor  decreases with cosmic time t in the past ($\dot{a}(t) < 0$) representing  the contracting phase of the universe, and the scale factor increases  with cosmic time t (i.e  $\dot{a}(t) > 0$)  representing the  expanding phase of the universe. Thus the scale factor attains a nonzero value at the transition point t=0, which implies that it is satisfying a bounce universe criterion.
In Fig. (2), three different phases of the evolution, where $H(t)<0$ for $t<0$ denotes contraction of the universe, $H(t)>0$ for $t>0$ denotes expansion of the universe, and $H(t)=0$ at $t=0$ denotes bouncing condition of the universe are found. Thus the universe contracted before the bounce and begins to expand after the bounce.  We plot the variation of the deceleration parameter with cosmic time for different values of $\sigma$ in Fig. (3), it is evident that as $\sigma$ increases, the duration of the bouncing epoch decreases.
\begin{figure}[H]
\centering
\includegraphics[width=8.41cm,height= 5.5 cm]{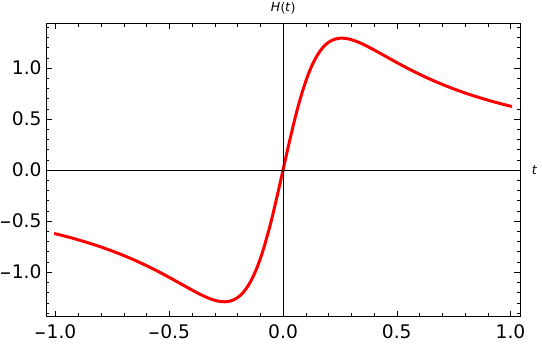}
\caption{Evolution of Scale factor versus cosmic time for $\sigma=10$ }
\end{figure}
\begin{figure}[H]
\centering
\includegraphics[width=8.41cm,height= 5.5 cm]{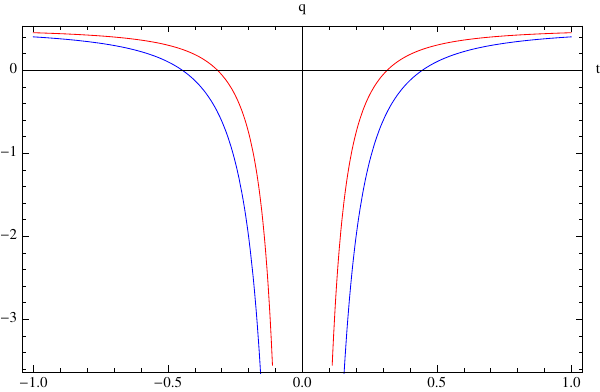}
\caption{Evolution of decleration parameter for $\sigma=20 $ (Red) and $\sigma=10$ (Blue)}
\end{figure}

We consider  modified gravity  $f(Q,C)$ as follows:
\begin{equation}
\label{e18}
f(Q,C)=  Q_0+\alpha Q+ \alpha_1 Q^2 + \beta C + \beta_1 C^2
\end{equation}
 where $\alpha,\; \alpha_1,\; \beta, \; \beta_1$  are the coupling parameter in the gravitational action with $Q_0$ the gravitational constant.
In the present case,  the  energy density, pressure, null energy condition (NEC), dominant energy condition (DEC) and strong energy condition (SEC) can  be expressed as,
\begin{equation}
\label{e19}
\rho=3\alpha H^2+\frac{Q_0}{2} -54 \alpha_1 H^4 - 18 \beta_1 (\dot{H}+3H^2)^2 +36 \beta_1 H (\ddot{H} +6H \dot{H}),
\end{equation} 

\begin{equation}
\label{e19a}
p= - 2\alpha \dot{H} -3\alpha H^2-\frac{Q_0}{2} +18 \alpha_1 H^2 ( 4 \dot{H} +3 H^2) +18 \beta_1 (\dot{H}+3H^2)^2 -12 \beta_1 \chi,
\end{equation} 

\begin{equation}
\label{e19b}
\rho+p = - 2 \alpha \dot{H}+ 72  \alpha_1 H^2 \dot{H} + 12  \beta_1 ( 18H^2 \dot{H} -\dddot{H} -3 H \ddot{H} -6 \dot{H}^2),
\end{equation}

\begin{equation}
\label{e19c}
\rho-p = Q_0+ 6\alpha H^2- 36 \alpha_1H^2(3 H^2+2 \dot{H})+12 \beta_1( \dddot{H}+ 6 \dot{H}^2+9 H \ddot{H}+18 \dot{H} H^2),
\end{equation}

\begin{equation}
\label{e19d}
\rho+3p =-Q_0- 6 \alpha H^2 - 6 \alpha \dot{H}+ 108\alpha_1 H^2 (H^2 +2 \dot{H}) +36 \beta_1 (9 H^4 -5 \dot{H}^2 -5 H \ddot{H}-\dddot{H}),
\end{equation}
where $\chi= 6 \dot{H}^2 +  6H \ddot{H} + \dddot{H} $
The equation of state parameter yields
\begin{equation}
\label{e19e}
\omega =\frac{ - 2\alpha \dot{H} -3\alpha H^2 +18 \alpha_1 H^2 ( 4 \dot{H} +3 H^2) +18 \beta_1 (\dot{H}+3H^2)^2 -12 \beta_1 \chi -\frac{Q_o}{2}}{3\alpha H^2-54 \alpha_1 H^4 - 18 \beta_1 (\dot{H}+3H^2)^2 +36 \beta_1 H (\ddot{H} +6H \dot{H}) +\frac{Q_0}{2}}
\end{equation}

\subsubsection{Case I}

 In this case we consider $\alpha=1$\; $\alpha_1=0$,\;$\beta_1=0$
then  $f(Q,C)= Q +\beta C +Q_o$, thereafter the equations (19), (20),(21),(22) and (23) can be rewritten as 
 \begin{equation}
 \label{e20}
     \rho= 3 H^2+\frac{Q_0}{2}
 \end{equation}
 \begin{equation}
 \label{e20a}
     p=-2\dot{H}-3 H^2- \frac{Q_0}{2}
 \end{equation}
 \begin{equation}
 \label{e20b}
     \rho+p= -2\dot{H}
 \end{equation}

\begin{equation}
\label{e20c}
    \rho-p= Q_0+6 H^2+2 \dot{H}
\end{equation}

\begin{equation}
\label{e20d}
    \rho+3 p= -Q_0 -6 \dot{H}-6 H^2
\end{equation}
 
 \begin{figure}
     \centering
     \begin{subfigure}[b]{0.4\textwidth}
         \centering
         \includegraphics[width=\textwidth]{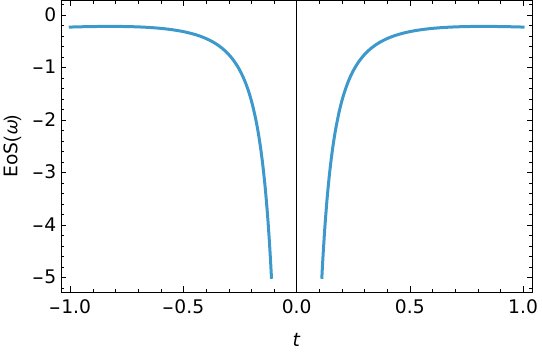}
         \caption{}
         \label{}
     \end{subfigure}
     \hfill
     \begin{subfigure}[b]{0.4\textwidth}
         \centering
         \includegraphics[width=\textwidth]{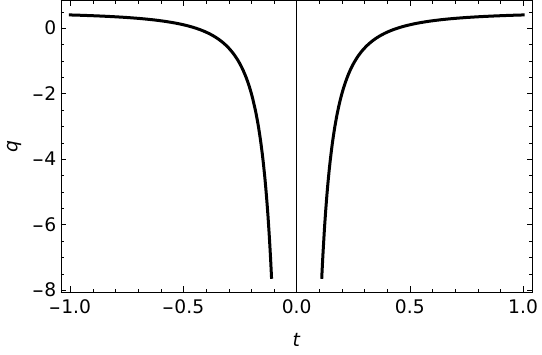}
         \caption{}
         \label{}
     \end{subfigure}
\caption{(a)Variation of EOS  parameter with cosmic time  for $\sigma=10$ ,\; $Q_o=0.5$,\;$\alpha_1=0$;\; $\beta_1=0$\\
        (b)Variation of deceleration parameter with cosmic time  for $\sigma=10$,\; $Q_o=0.5$,\;$\alpha_1=0$;\; $\beta_1=0$}
        \label{}
\end{figure}

We plot the variation of EoS parameter against the cosmic time in Fig. 4(a), it indicates that during and after the bounce, the universe is described by phantom fluid which however transits to quintessence. There is a transition from accelerating universe to decelerating universe (Fig. 4(b)).  \\

\begin{figure}
     \centering
     \begin{subfigure}[b]{0.4\textwidth}
         \centering
         \includegraphics[width=\textwidth]{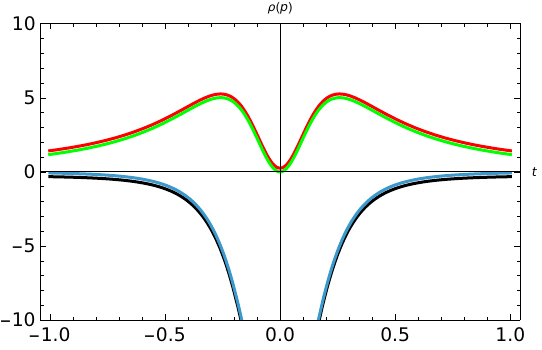}
         \caption{}
         \label{}
     \end{subfigure}
     \hfill
     \begin{subfigure}[b]{0.4\textwidth}
         \centering
         \includegraphics[width=\textwidth]{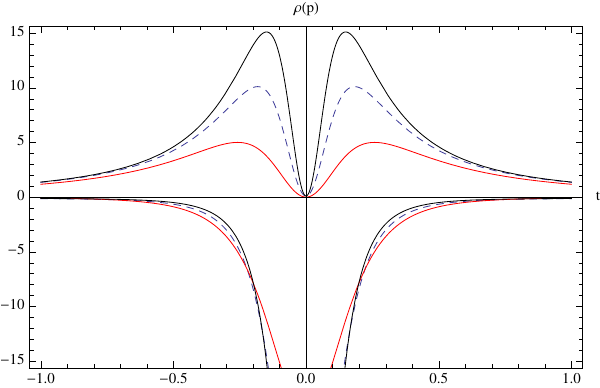}
         \caption{}
         \label{}
     \end{subfigure}%
     \hfill
     \begin{subfigure}[b]{0.4\textwidth}
         \centering
         \includegraphics[width=\textwidth]{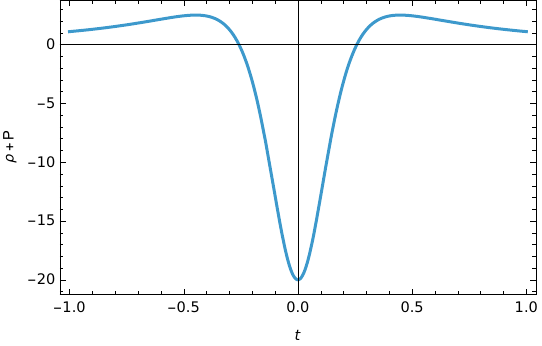}
         \caption{}
         \label{}
     \end{subfigure}
     \hfill
     \begin{subfigure}[b]{0.4\textwidth}
         \centering
         \includegraphics[width=\textwidth]{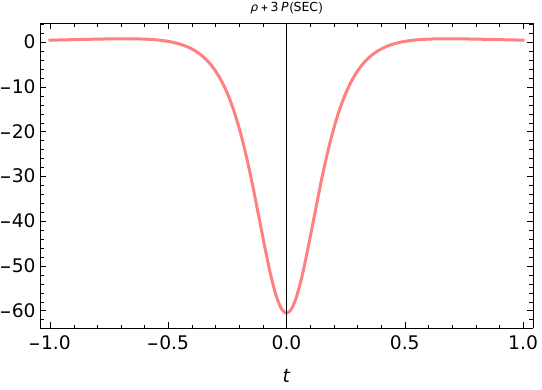}
         \caption{}
         \label{}
     \end{subfigure}
        \caption{(a)variation of Energy density with cosmic time for  $\sigma=10$  $Q_0$ = 0 (Green) and  $Q_0$ = 0.5 (Red) or variation of pressure with cosmic time  for $Q_0$ = 0 (Blue) and   $Q_0$ = 0.5 (Black)\\
        (b)Variation of Energy density and pressure with cosmic time  for $\sigma=10 $ (Red), 20 (Dashed) and 30 (Black) without $Q_o$\\
        (c)variation of NEC ($\rho+p$)with cosmic time(blue),\;and \\
        (d)variation of SEC ($\rho+3p$)(purple) with cosmic time for   $Q_0$ = 0.5 ,\;$\alpha_1=0$;\; $\beta_1=0$}
        \label{}
\end{figure}

The  energy density/pressure are plotted in Figs.5 (a) and (b) with cosmic time shows that energy density is positive and the pressure is negative near the bounce.  We also noted that as $\sigma $ increases the peak of the energy density near the bounce increases. NEC  and SEC are plotted in Fig 5(c) and (d) it is found that both NEC and SEC are violated near the bounce but NEC is found to obey away from the bounce.  We note that there is no effect of $\beta$ on the energy density and pressure, thus the bounce is independent of the boundary term. \\

 \subsubsection{Case II: }
 In this case  we consider modified gravity $f(Q,C)$ with $\alpha=1$\; $\alpha_1 \ne 0$,\;$\beta_1 \ne 0$, which follows:
 $f(Q,C)= \alpha Q +\alpha_1 Q^2 +\beta_1 C^2+Q_0$.
The equations (19), (20), (21), (22), and (23) for the energy density, pressure, NEC, DEC, and SEC can be expressed as 
 
\begin{equation}
\label{e21}
\rho=3 H^2+\frac{Q_0}{2} +54\alpha_1 H^4 - 18 \beta_1 (\dot{H}+3H^2)^2 +36 \beta_1 H (\ddot{H} +6H \dot{H})
\end{equation} 

\begin{equation}
\label{e21a}
p= - 2\dot{H} -3 H^2-\frac{Q_o}{2} -18 \alpha_1 H^2 ( 4 \dot{H} +3 H^2) +18 \beta_1 (\dot{H}+3H^2)^2 -12 \beta_1 (6 \dot{H}^2 +  6H \ddot{H} + \dddot{H} )
\end{equation} 

\begin{equation}
\label{e21b}
\rho+p = - 2 \dot{H}- 72 H^2 \dot{H} + 12  \beta_1 ( 18H^2 \dot{H} -\dddot{H} -3 H \ddot{H} -6 \dot{H}^2)
\end{equation} 

 \begin{equation}
 \label{e21c}
\rho-p = Q_0+ 6 H^2+ 36 \alpha_1H^2(3 H^2+2 \dot{H})+12 \beta_1( \dddot{H}+ 6 \dot{H}^2+9 H \ddot{H}+18 \dot{H} H^2)
\end{equation}

\begin{equation}
\label{e21d}
\rho+3p =-Q_0- 6 H^2 - 6\dot{H}- 108\alpha_1 H^2 (H^2 +2 \dot{H}) +36 \beta_1 (9 H^4 -5 \dot{H}^2 -5 H \ddot{H}-\dddot{H})
\end{equation}

\begin{figure}
     \centering
     \begin{subfigure}[b]{0.4\textwidth}
         \centering
         \includegraphics[width=\textwidth]{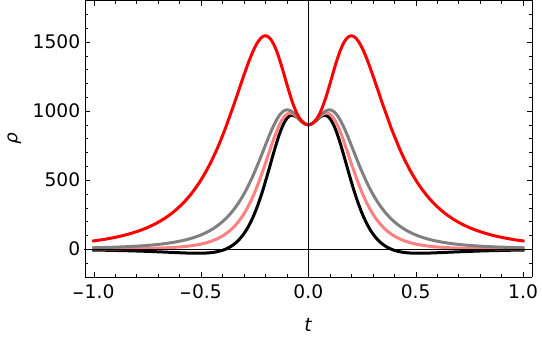}
         \caption{}
         \label{}
     \end{subfigure}
     \hfill
     \begin{subfigure}[b]{0.4\textwidth}
         \centering
         \includegraphics[width=\textwidth]{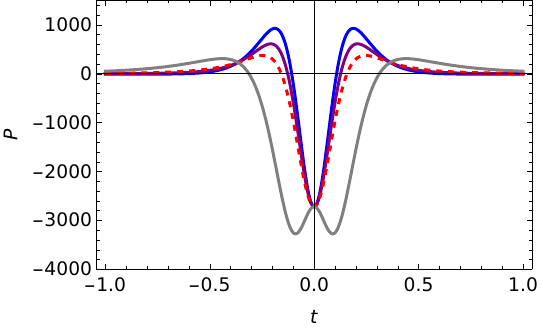}
         \caption{}
         \label{}
     \end{subfigure}%
     \hfill
     \begin{subfigure}[b]{0.4\textwidth}
         \centering
         \includegraphics[width=\textwidth]{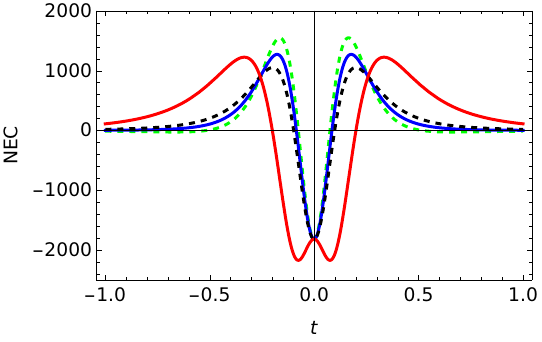}
         \caption{}
         \label{}
     \end{subfigure}
     \hfill
     \begin{subfigure}[b]{0.4\textwidth}
         \centering
         \includegraphics[width=\textwidth]{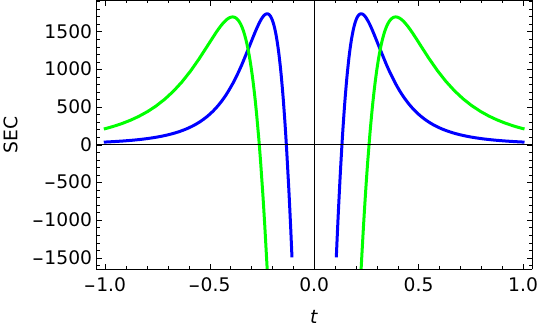}
         \caption{}
         \label{}
     \end{subfigure}
        \caption{(a)Variation of Energy density   with cosmic time  for $\sigma=10$,\;$Q_o=0.5$,\;$\beta_1=-0.5$,\;$\alpha=1$,\;$\alpha_1=-5$(red);\;and$\alpha_1=1$(Gray),\;$\alpha_1=2$(pink),\;$\alpha_1=3$(Black)\\
        (b)Variation of Pressure   with cosmic time  for $\sigma=10$,\;$Q_o=0.5$,\;$\beta_1=-0.5$,\;$\alpha=1$,\;$\alpha_1=-5$(Gray);\;and$\alpha_1=1$(Red dotted),\;$\alpha_1=2$(purple),\;$\alpha_1=3$(Blue)\\
        (c)Variation of NEC  with cosmic time  for $\sigma=10$,\;$Q_o=0.5$,\;$\beta_1=-0.5$,\;$\alpha=1$,\;$\alpha_1=-5$(red);\;and$\alpha_1=1$(Black dotted),\;$\alpha_1=2$(Blue),\;$\alpha_1=3$(Green dotted)\\
            (d)Variation of SEC for $\sigma=10 $,\;  $\beta_1=-0.5$\;$\alpha=1$,\;$\alpha_1=-5$(Green);\;$\alpha_1=1$(Blue)} 
        \label{}
\end{figure}

In Fig 6(a) as $\alpha_1$ increases we find that at t=0 energy density is same but away from it the bounce decreases as $\alpha_1$ increases and $\beta_1 < 0$.However we note that for$\alpha_1\geq 3$ the energy density though positive near the bounce decreases as time increases but subsequently it attains a negative value which is not realistic.

\begin{figure}
     \centering
     \begin{subfigure}[b]{0.4\textwidth}
         \centering
         \includegraphics[width=\textwidth]{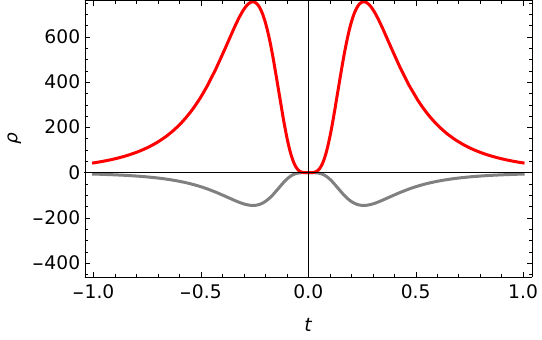}
         \caption{}
         \label{}
     \end{subfigure}
     \hfill
     \begin{subfigure}[b]{0.4\textwidth}
         \centering
         \includegraphics[width=\textwidth]{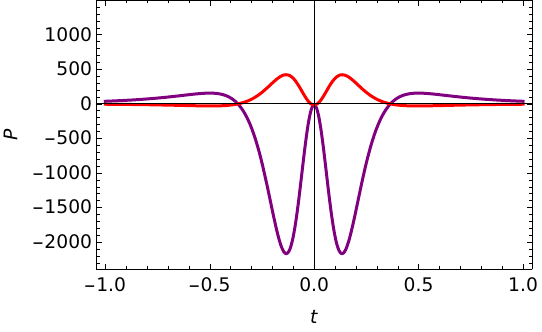}
         \caption{}
         \label{}
     \end{subfigure}%
     \hfill
     \begin{subfigure}[b]{0.4\textwidth}
         \centering
         \includegraphics[width=\textwidth]{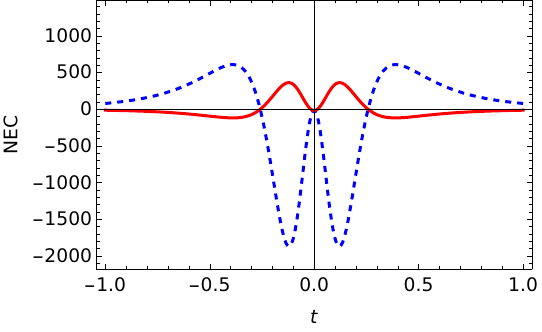}
         \caption{}
         \label{}
     \end{subfigure}
     \hfill
     \begin{subfigure}[b]{0.4\textwidth}
         \centering
         \includegraphics[width=\textwidth]{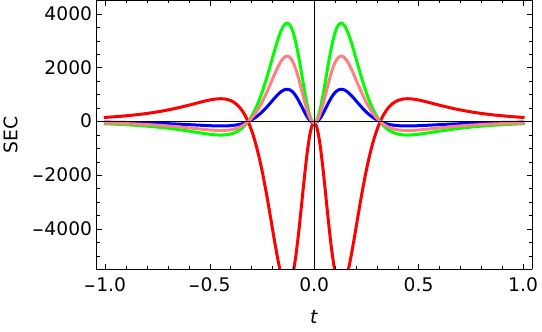}
         \caption{}
         \label{}
     \end{subfigure}
        \caption{(a)Variation of Energy density   with cosmic time  for $\sigma=10$,\;$Q_o=0.5$,\;$\beta_1=0.0005$,\;$\alpha=1$,\;$\alpha_1=-5$(red);\;and\;$\alpha_1=1$(Gray)\\
        (b) Variation of Pressure   with cosmic time  for $\sigma=10$,\;$Q_o=0.5$,\;$\beta_1=0.0005$,\;$\alpha=1$,\;$\alpha_1=-5$(purple);\;and$\alpha_1=1$(Red)\\
        (c)Variation of NEC   with cosmic time  for $\sigma=10$,\;$Q_o=0.5$,\;$\beta_1=0.0005$,\;$\alpha=1$,\;$\alpha_1=-5$(Blue dotted);\;and$\alpha_1=1$(Red)\\
        (d)Variation of SEC for $\sigma=10 $,\;  $\beta_1=0.0005$\;$\alpha=1$,\;$\alpha_1=-5$(Red dotted);\;$\alpha_1=1$(Blue ),\;$\alpha_1=2$(Pink),\;$\alpha_1=3$(Green)}
        \label{}
\end{figure}

In Fig 6(b) pressure is negative at the bounce having two maxima with $p<0$ and $p>0$ for $\alpha_1= -5$ but for others values we get one minima and one maxima for  $\alpha_1>0$. In Figs. 6(c) and (d) we plot NEC and SEC which show that near the bounce both energy conditions are violated but away from the bounce they obey.

In Fig 7(a) it is found that energy density is positive for $\alpha_1<0 $ negative for $\alpha_1>0 $. It is evident from Figs. 7(a) and (b) that at the bounce $\rho=0 $ and $p=0$ thus bounce may begin/end in vaccum. It is evident that both NEC and SEC are obeyed for $\alpha_1>0$.

We plot variation of the energy density in Figs. 6(a) and 7(a), it is evident that energy density is positive definite, minimum at the bounce thereafter increasing attains a  maximum which then  decreases for a set of non-zero values of $\alpha$,  $\beta$, $\beta_1$, $\alpha_1$, and $Q_0$ values.  The pressure is  negative near the bounce with a maximum which decreases thereafter attains minimum it increases attains a maximum and then after a specific time depending upon the model parameters it permits a universe with positive pressure attains a maximum and thereafter decreases slowly towards a vanishing value.  There is a flip of sign of the pressure  from negative to positive value and then $p \rightarrow 0$.
 
\begin{figure}[H]
\centering
\includegraphics[width=8.41cm,height= 5.5 cm]{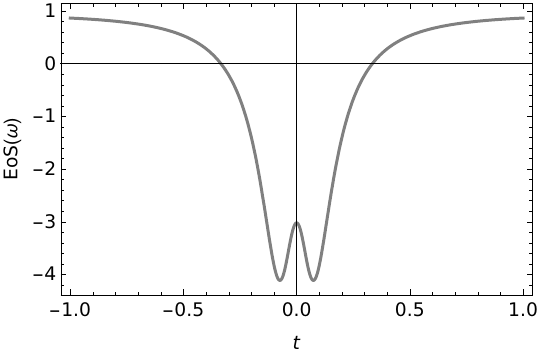}
\caption{ Variation of EoS with cosmic time  for $\sigma=10$ ,\; $Q_o=0.5$,\;$\alpha_1=-10$;\; $\beta_1=-0.5$}
\end{figure}

The variation of the EoS parameter with time is shown in Fig. (8), it is evident that during the bounce and over a period  $\omega< -1$ (phantom region). Here, the behavior of the EoS parameter is identical to that obtained in the symmetric bounce model. This means that the physical conditions governing the cosmic behavior are stable and do not reach infinite values at any point during the bounce, ensuring a smooth transition through the critical phase. We note a different picture from that obtained in Fig. 4(a), where one gets single bounce, but double bounce is found in Fig.(8).

The variation of the deceleration parameter ($q$) plotted in Fig.(9) shows that an accelerating universe transits into accelerating universe.

\begin{figure}[H]
\centering
\includegraphics[width=8.41cm,height= 5.5 cm]{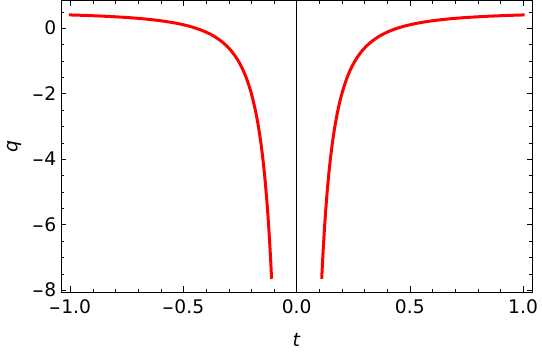}
\caption{Variation of deceleration parameter ($q$)  with cosmic time(t)  for $\sigma=10$ ,\; $Q_o=0.5$,\;$\alpha_1=-10$;\; $\beta_1=-0.5$}
\end{figure}

In Fig 6(a) it is evident that for all values of $\alpha_1$ and $\beta_1<0$ energy density is always positive and there is a bounce near t=0, pressure($p$) is decreasing with a certain negative value, then increasing near the bounce, again it is decreasing. It is also shown that both NEC and SEC are violated here.

\subsection{Model II : Emergent Universe}

 In this section, we consider an emergent universe with  scale factor given by
 \begin{equation}
 \label{e22}
a(t)= a_o  e^{\zeta t^2}.
\end{equation}
 where $a_0$ and $\zeta$ are arbitrary constants that satisfy the conditions to obtain a bouncing universe for the scale factor. 
 \begin{equation}
 \label{e22a}
H(t) = 2 \zeta t.
\end{equation}
The above scale factor ensures a bouncing scenario at $t=t_c=0$, which satisfies the bouncing condition $H<0$ for $t<t_c$, $H>0$ for $t> t_c$, and $H=0$ at $t=t_c$. At the bouncing point, one obtains $\dot{H}= 2\zeta$ greater than zero. 
The deceleration parameter for the bouncing scenario is  given by
 \begin{equation}
\label{e23}
q(t)= -1+ \frac{d}{dt} \left(H^{-1}\right) = -1 - \frac{1}{2 \zeta \; t^2}. 
\end{equation}

\begin{figure}[H]
\centering
\includegraphics[width=8.41cm,height= 5.5 cm]{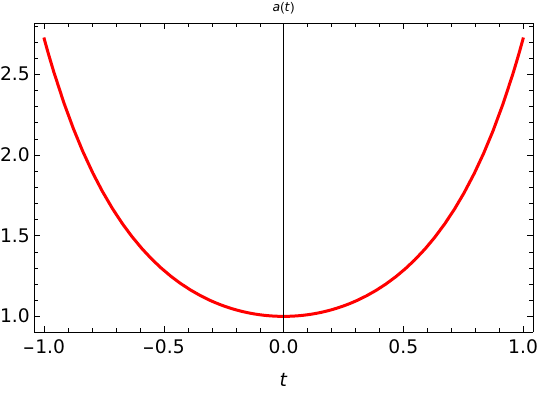}
\caption{Evolution of Scale factor versus cosmic time  for $\zeta=1 $ }
\end{figure}

\begin{figure}[H]
\centering
\includegraphics[width=8.41cm,height= 5.5 cm]{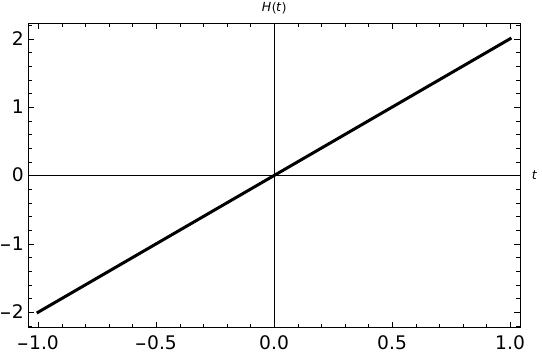}
\caption{Evolution of Hubble parameter versus cosmic time for $\zeta=1$ }
\end{figure}

In Fig. (10), it is evident that a contracting phase of the universe is permitted with no singularity when the scale factor is a decreasing with time and an expanding phase of the universe when the scale factor is an increasing function of cosmic time t,  scale factor is a non zero at the transition point ($t=0$),  thus the scale factor permits a bounce cosmology scenario.
In Fig. (11), three different phases of the universe, where $H(t)<0$ for $t<0$ denotes contraction of the universe, $H(t)>0$ for $t>0$ denotes expansion of the universe, and $H(t)=0$ at $t=0$ denotes bouncing condition of the universe. The present cosmological model is found to contract before bouncing and expansion after the bounce.
 In Fig.  (12), we plot deceleration parameter with cosmic time, an accelerating phase of the  universe is  permitted with  a $q>-1$ transits to a universe with $q=-1$ after the bounce.

\begin{figure}[H]
\centering
\includegraphics[width=8.41cm,height= 5.5 cm]{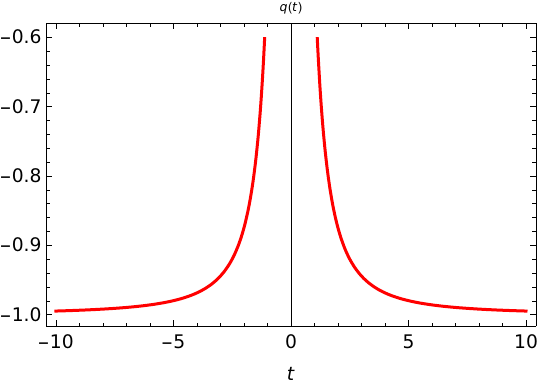}
\caption{Evolution of decleration parameter for $\zeta=1 $ }
\end{figure}

\subsubsection{Case 1:}

 We consider $\alpha=1$\; $\alpha_1=0$,\;$\beta_1=0$ and the modified gravity is $f(Q,C)= Q +\beta C +Q_0$.
Eqs. (19), (20), and (21) become
 \begin{equation}
 \label{e24}
     \rho= 3 H^2+\frac{Q_0}{2}
 \end{equation}
 \label{e24a}
 \begin{equation}
     p=-2\dot{H}-3 H^2- \frac{Q_0}{2}
 \end{equation}
 \begin{equation}
 \label{e24b}
     \rho+p= -2\dot{H}
 \end{equation}

In Fig.  13(a), the plot of energy density for $\zeta=1 $,\; $\alpha_1=0$,\; $\beta_1=0$,\; $Q_0=1.5$ eis found positive throughout and the behaviour is independent of   boundary term $C$.
In Fig. 13(b), we plot pressure for the same set of values which is always negative and the NEC plotted in Fig. 13(c) is  found to violate always during evolution.
  SEC is violated as shown in the plots Figs. 13(d). 
 \begin{figure}
     \centering
     \begin{subfigure}[b]{0.4\textwidth}
         \centering
         \includegraphics[width=\textwidth]{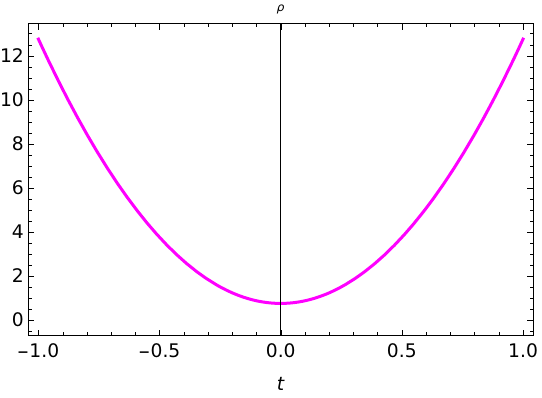}
         \caption{}
         \label{}
     \end{subfigure}
     \hfill
     \begin{subfigure}[b]{0.4\textwidth}
         \centering
         \includegraphics[width=\textwidth]{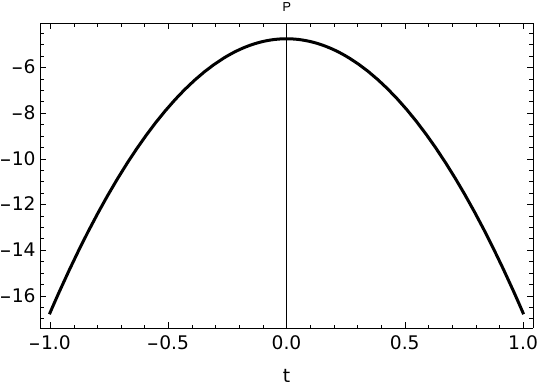}
         \caption{}
         \label{}
     \end{subfigure}%
     \hfill
     \begin{subfigure}[b]{0.4\textwidth}
         \centering
         \includegraphics[width=\textwidth]{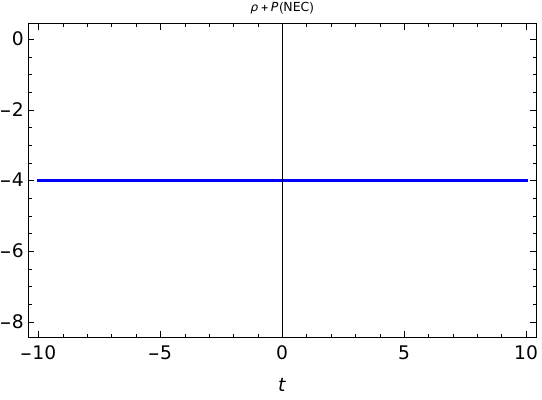}
         \caption{}
         \label{}
     \end{subfigure}
     \hfill
     \begin{subfigure}[b]{0.4\textwidth}
         \centering
         \includegraphics[width=\textwidth]{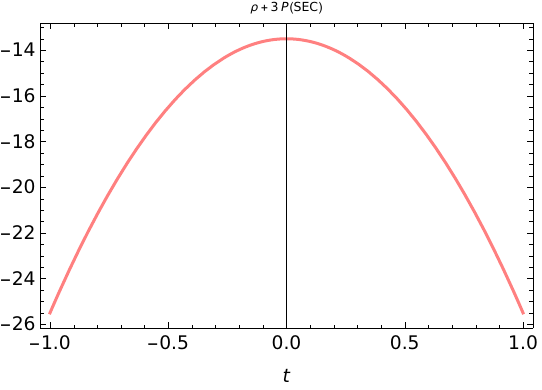}
         \caption{}
         \label{}
     \end{subfigure}
        \caption{(a)Variation of Energy density with cosmic time for $\zeta=1 $,\; $\alpha_1=0$,\; $\beta_1=0$,\; $Q_0=1.5$\\
        (b)Variation of Pressure with cosmic time for $\zeta=1 $,\; $\alpha_1=0$,\; $\beta_1=0$,\; $Q_0=1.5$ 
        (c) NEC for $\zeta=1 $ ,\; $\alpha_1=0$,\; $\beta_1=0$,\; $Q_0=1.5$\\
        (d) SEC for $\zeta=1 $ ,\; $\alpha_1=0$,\; $\beta_1=0$,\; $Q_0=1.5$}
         \label{}
\end{figure}

 \subsubsection{Case 2:}
 In this case we consider $\alpha=1$,\;$\alpha_1 \ne 0$,\;$\beta_1\ne0$ and the corresponding modified gravity becomes
 $f(Q,C)= Q +\alpha_1 Q^2+\beta C +\beta_1 C^2+Q_0$.
 Equations. (19), (20), and (21) can be expressed as 
 \begin{equation}
 \label{e25}
     \rho=3\alpha H^2+\frac{Q_0}{2}-54\alpha_1H^4 - 18 \beta_1 (\dot{H}+3H^2)^2 +36 \beta_1 H (\ddot{H} +6H \dot{H})
 \end{equation}
 \label{e25a}
 \begin{equation}
p= - 2\alpha \dot{H}-3\alpha H^2 +18\alpha_1H^2(4\dot{H}+3 H^2)-\frac{Q_0}{2}+18 \beta_1 (\dot{H}+3H^2)^2 -12 \beta_1 (6 \dot{H}^2 +  6H \ddot{H} + \dddot{H} )
\end{equation} 

\begin{equation}
\label{e25b}
\rho+p = - 2 \alpha \dot{H}+72\alpha_1 \dot{H} H^2 + 12  \beta_1 ( 18H^2 \dot{H} -\dddot{H} -3 H \ddot{H} -6 \dot{H}^2)
\end{equation} 

\begin{figure}[H]
\centering
\includegraphics[width=8.41cm,height= 5.5 cm]{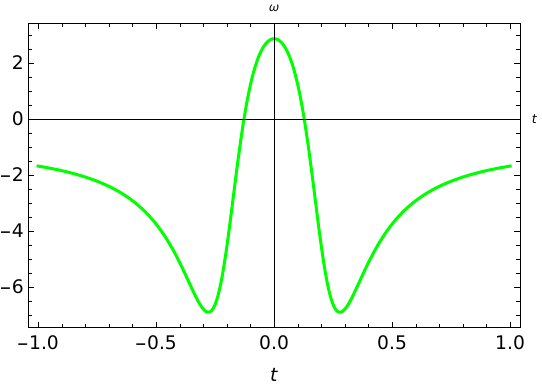}
\caption{Evolution of EoS parameter with cosmic time for $\zeta=1 $,\; $\alpha_1=-10$,\; $\beta_1=-0.5$,\; $Q_0=0.5$ }
\end{figure}
 The evolution of the equation of state parameter with cosmic time is shown in Fig. (14), there is a  single bounce in the universe near $\omega >0$ but eventually it decreases attains a minimum thereafter increases but remains negative $\omega <-1$ (Phantom region).

 \begin{figure}
     \centering
     \begin{subfigure}[b]{0.4\textwidth}
         \centering
         \includegraphics[width=\textwidth]{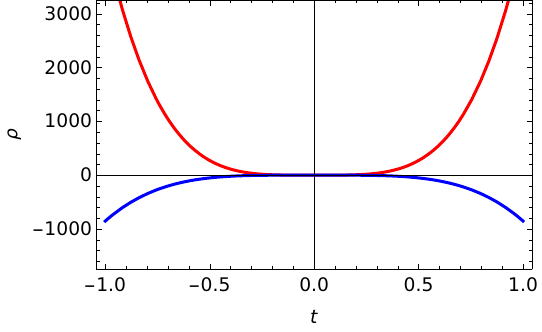}
         \caption{}
         \label{}
     \end{subfigure}
     \hfill
     \begin{subfigure}[b]{0.4\textwidth}
         \centering
         \includegraphics[width=\textwidth]{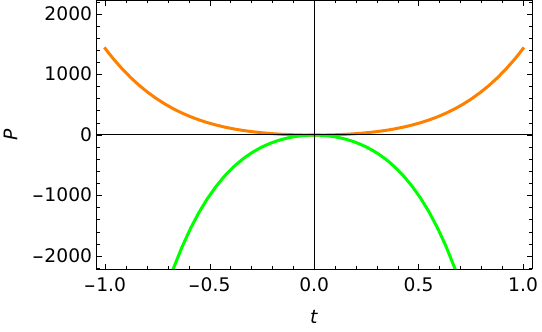}
         \caption{}
         \label{}
     \end{subfigure}%
     \hfill
     \begin{subfigure}[b]{0.4\textwidth}
         \centering
         \includegraphics[width=\textwidth]{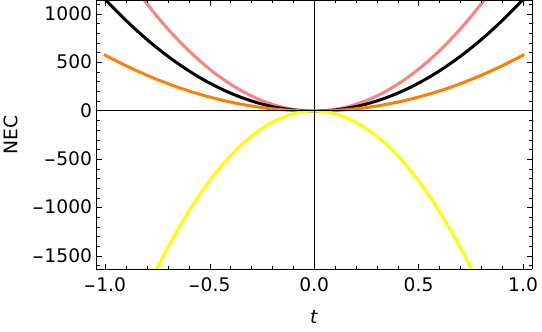}
         \caption{}
         \label{}
     \end{subfigure}
     \hfill
     \begin{subfigure}[b]{0.4\textwidth}
         \centering
         \includegraphics[width=\textwidth]{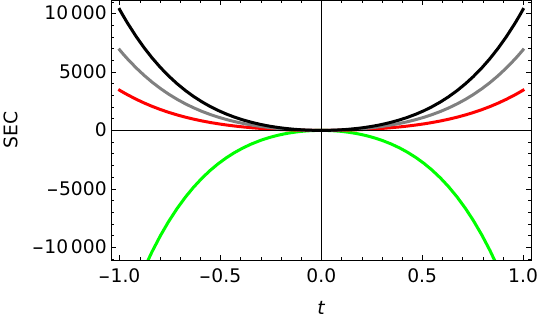}
         \caption{}
         \label{}
     \end{subfigure}
        \caption{(a)variation of Energy density with cosmic time for $\zeta=1 $,\;$\beta_1=0.0005$,\; $Q_0=0.5$,\; $\alpha_1=-5$(Red ),\;$\alpha_1=1$(Blue)  \\
        (b)variation of Pressure with cosmic time for $\zeta=1 $,\;$\beta_1=0.0005$,\; $Q_0=0.5$,\; $\alpha_1=-5$(green ),\;$\alpha_1=1$(orange) \\
        (c) variation of NEC with cosmic time for $\zeta=1 $,\;$\beta_1=0.0005$,\; $Q_0=0.5$,\; $\alpha_1=-5$(yellow),\;$\alpha_1=1$(orange),\;$\alpha_1=2$(black),\;$\alpha_1=3$(pink) \\
        (d) Variation of SEC for $\zeta=1 $ ,\;$\beta_1=0.0005$,\; $Q_0=0.5$,\; $\alpha_1=-5$(green),\;$\alpha_1=1$(red),\;$\alpha_1=2$(gray),\;$\alpha_1=3$(black)  }
         \label{}
\end{figure}
 In Fig 15(a) it is evident that for $\zeta=1$,\;$\alpha_1<0$ , energy density $\rho>0$ and for $\alpha_1>0$ energy density is negative, the later case is unrealistic.\\
In Fig 15(b) it is evident that for $\zeta=1$,\;$\alpha_1<0$, and $\beta>0$, pressure $p<0$, and for $\alpha_1>0$ pressure is positive.\\
In Figs. 15(c) and (d) it is evident that for $\zeta=1$,\;$\alpha_1<0$, and $\beta>0$, NEC  and SEC are violated here but for  positive $\alpha_1$  that is $\alpha_1>0$  both NEC, SEC  are always obeyed which is new in the model. \\

\begin{figure}
     \centering
     \begin{subfigure}[b]{0.4\textwidth}
         \centering
         \includegraphics[width=\textwidth]{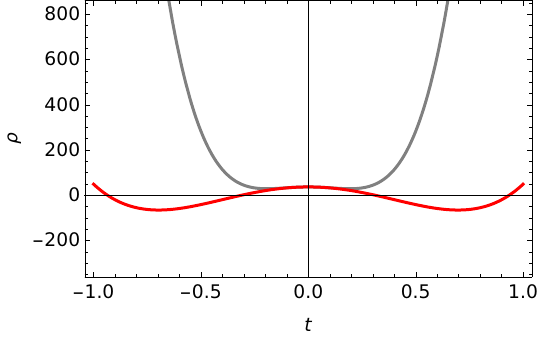}
         \caption{}
         \label{}
     \end{subfigure}
     \hfill
     \begin{subfigure}[b]{0.4\textwidth}
         \centering
         \includegraphics[width=\textwidth]{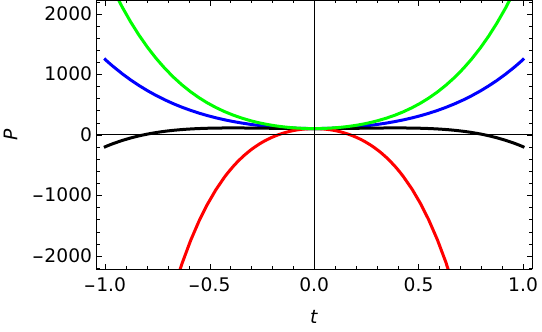}
         \caption{}
         \label{}
     \end{subfigure}%
     \hfill
     \begin{subfigure}[b]{0.4\textwidth}
         \centering
         \includegraphics[width=\textwidth]{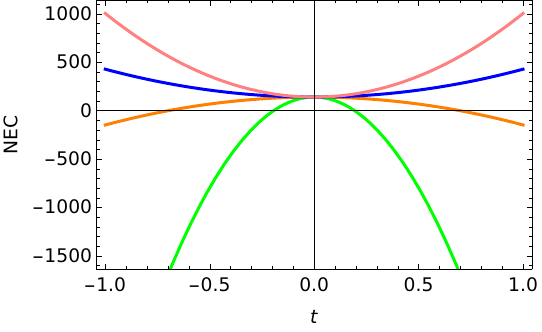}
         \caption{}
         \label{}
     \end{subfigure}
     \hfill
     \begin{subfigure}[b]{0.4\textwidth}
         \centering
         \includegraphics[width=\textwidth]{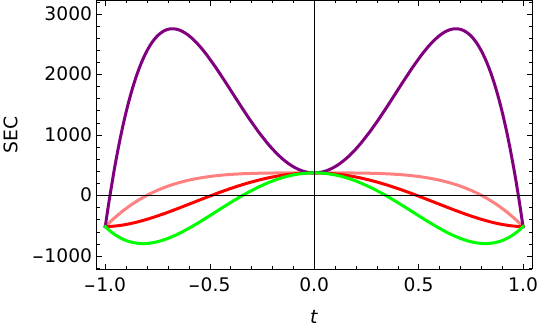}
         \caption{}
         \label{}
     \end{subfigure}
        \caption{(a)variation of Energy density with cosmic time for $\zeta=1 $,\;$\beta_1=-0.5$,\; $Q_0=0.5$,\; \;$\alpha_1=-5$(gray),\;$\alpha_1=1$(red) \\
        (b) variation of pressure with cosmic time for $\zeta=1 $,\;$\beta_1=-0.5$,\; $Q_0=0.5$,\; \;$\alpha_1=-5$(red),\;$\alpha_1=1$(black),\;$\alpha_1=2$(blue),\;$\alpha_1=1$(green)\\
        (c)variation of NEC with cosmic time for $\zeta=-1 $,\;$\beta_1=-0.5$,\; $Q_0=0.5$,\; $\alpha_1=-5$(Green),\;$\alpha_1=1$(orange),\;$\alpha_1=2$(blue),\;$\alpha_1=3$(pink) \\
        (d)variation of SEC with cosmic time for $\zeta=-1 $,\;$\beta_1=-0.5$,\; $Q_0=0.5$,\; $\alpha_1=-5$(purple),\;$\alpha_1=1$(pink),\;$\alpha_1=2$(red),\;$\alpha_1=3$(Green)}
         \label{}
\end{figure}
 
In Fig 16(a) it is evident that for $\zeta=1$, \; $\alpha_1<2$ \;  we obtain positive energy density ($\rho>0$) always near the bounce. In Fig 16(b) we see that $\alpha_1<0$ $p$ is negative and for $\alpha_1>0$ $p$ is positive here.NEC is satisfied for $\alpha_1>0$ which is shown in Fig 16(c), for $\alpha_1<0$ NEC is violated. SEC is obeyed for any value of $\alpha_1$ which is shown in Fig 16(d).

\subsection{Exponential $f(Q,C)$ gravity}

In this section we consider an exponential form of $f(Q,C)$ gravity  \cite{77} given by: 
\begin{equation}
\label{e26}
f(Q,C)=  -Qe^{\frac{\alpha Q_0}{Q}}+ \beta C + \beta_1 C^2
\end{equation}
 where $\alpha,\; \beta, \; \beta_1$ are the gravitational action coupling constant and $Q_0$ is an arbitrary constant. 
Here the energy density, pressure, and NEC are given by 
\begin{equation}
\label{e27}
\rho=-(3 H^2+\alpha Q_o)e^{\frac{-\alpha Q_0}{6 H^2}}  - 18 \beta_1 ( \dot{H}^2+9 H^4-6H^2 \dot{H}  -2 H \dddot{H})
\end{equation} 
\[
p= (3 H^2 +\alpha Q_0 +2 \dot{H}+ \frac{\alpha Q_0 \dot{H}}{3 H^2} ) e^{\frac{-\alpha Q_0}{6 H^2}}+\frac{\alpha ^2 Q_0^2 \dot{H}}{9 H^4}e^{\frac{-\alpha Q_0}{6 H^2}} \]
\begin{equation}
\label{e27a}
\;\;\;\;\; \;\;\; - 18\beta_1(3 \dot{H}^2-9 H^4-6 \dot{H}H^2)-12 \beta_1(\dddot{H}+6 H \ddot{H})
\end{equation}

\begin{figure}
     \centering
     \begin{subfigure}[b]{0.4\textwidth}
         \centering
         \includegraphics[width=\textwidth]{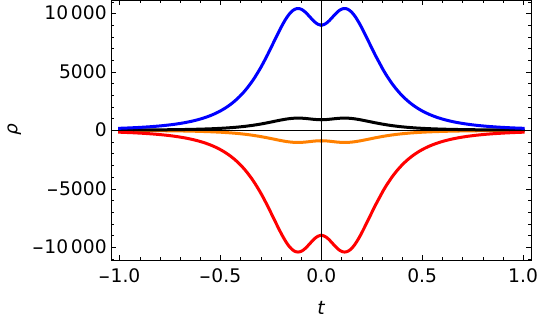}
         \caption{}
         \label{}
     \end{subfigure}
     \hfill
     \begin{subfigure}[b]{0.4\textwidth}
         \centering
         \includegraphics[width=\textwidth]{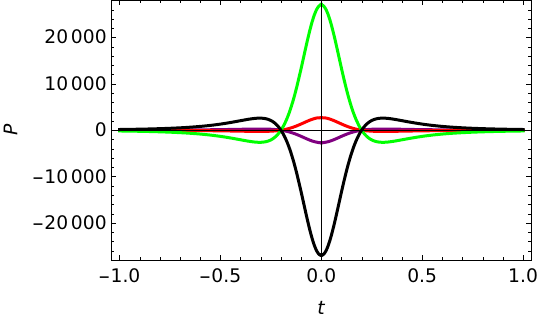}
         \caption{}
         \label{}
     \end{subfigure}%
     \hfill
     \begin{subfigure}[b]{0.4\textwidth}
         \centering
         \includegraphics[width=\textwidth]{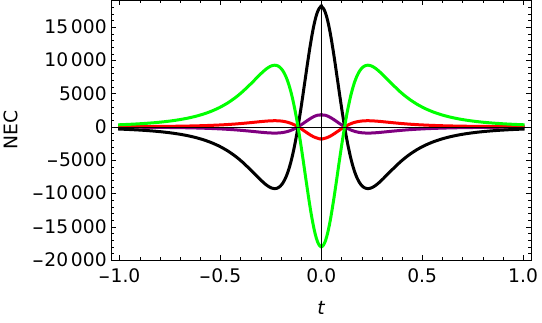}
         \caption{}
         \label{}
     \end{subfigure}
     \hfill
     \begin{subfigure}[b]{0.4\textwidth}
         \centering
         \includegraphics[width=\textwidth]{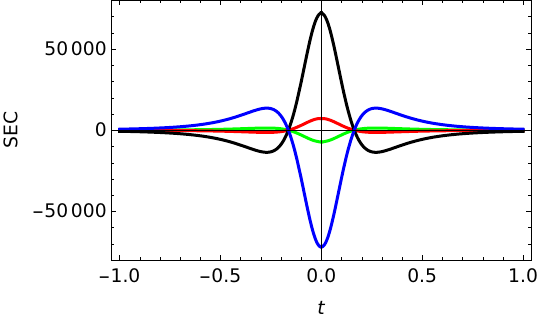}
         \caption{}
         \label{}
     \end{subfigure}
        \caption{(a)Variation of Energy density with cosmic time for $\sigma=10 $,\; $Q_0=0.5$, \; $\alpha=1$,\; $\beta_1=0.5$(orange),\;$\beta_1=-0.5$(black) and $\alpha=-0.0005$,\;$\beta_1=5$(red),\;$\beta_1=-5$(Blue) in exponential modified gravity\\
        (b)Variation of pressure with cosmic time for $\sigma=10 $,\; $Q_0=0.5$, \; $\alpha=1$,\; $\beta_1=-0.5$(Purple),\;$\beta_1=0.5$(red) and $\alpha=-0.0005$,\;$\beta_1=5$(green),\;$\beta_1=-5$(black) in exponential modified gravity \\
        (c)Variation of NEC with cosmic time for $\sigma=10 $,\; $Q_0=0.5$, \; $\alpha=1$,\; $\beta_1=-0.5$(red),\;$\beta_1=0.5$(purple) and $\alpha=-0.0005$,\;$\beta_1=5$(black),\;$\beta_1=-5$(blue) in exponential modified gravity\\
        (d)Variation of SEC with cosmic time for $\sigma=10 $,\; $Q_0=0.5$, \; $\alpha=1$,\; $\beta_1=-0.5$(Green),\;$\beta_1=0.5$(red) and $\alpha=-0.0005$,\;$\beta_1=5$(black),\;$\beta_1=-5$(blue) in exponential modified gravity  }
         \label{}
\end{figure}

\[    
\rho+p = \left[\frac{\alpha Q_0 \dot{H}}{3 H^2}+ \left (\frac{\alpha^2 Q_0 ^2\dot{H}}{9 H^4}\right)\right]e^{\frac{-\alpha Q_0}{6 H^2}}+2\dot{H}e^{\frac{-\alpha Q_0}{6 H^2}}
\]
\begin{equation}
\label{e27b}
\;\;\;-12\beta_1(\dddot{H}+3 H\ddot{H}-18\dot{H}H^2+6 \dot{H}^2)
\end{equation}

\begin{equation}
\label{e28}
\omega =\frac{\left(3 H^2 +2 \dot{H}+\alpha Q_0 + \frac{\alpha Q_0 \dot{H}}{3 H^2} \right) e^{\frac{-\alpha Q_0}{6 H^2}}+ \frac{\alpha ^2 Q_0^2 \dot{H}}{9 H^4} e^{\frac{-\alpha Q_0}{6 H^2}}- 6 \beta_1 \xi}{-(3 H^2+ Q_o)e^{\frac{-\alpha Q_0}{6 H^2}} -18 \beta_1 \xi_1}
\end{equation} 
where $\xi= 9 \dot{H}^2-27H^4-18\dot{H}H^2+2\dddot{H}+12H\ddot{H}$, $\xi_1=    \dot{H}^2+9H^4-6\dot{H}H^2-2 H \dddot{H}  $.\\

For power law given by eq. (15), we plot energy density in Fig.17(a), we note that $\rho>0 $ for $\beta_1<0$ with $\alpha =1$ but for $\alpha<0$ it is also satisfied $\rho>0$ for $\beta_1<0$. It is found that the energy density is positive, and there is a single bounce at t = 0. The variation of pressure in Fig. 17(b) for $\alpha>0$ and $\beta_1<0$, we note that $p<0$ but in case of $\alpha<0$ and $\beta_1=5$ we note that $p>0$  indicates that pressure is negative around the bounce point, but it increases from a minimum to attain a maximum and then decreases.\\

In Fig 17(c) NEC is obeyed for $\alpha>0$ and $\beta_1<0$ and also for $\alpha<0$ and $\beta_1<0$, but NEC is violated for $\alpha>0$ and  $\beta_1>0$ and also for $\alpha<0$ and  $\beta_1>0$.\\

In Fig 17(d) SEC is satisfied for $\alpha <0$ and $\beta_1<0$ and also for $\alpha >0$ and $\beta_1=-5$ but it is violated for $\alpha >0$ and $\beta_1>0$ and also for $\alpha <0$ and $\beta_1>0$.

\begin{figure}[H]
\centering
\includegraphics[width=8.41cm,height= 5.5 cm]{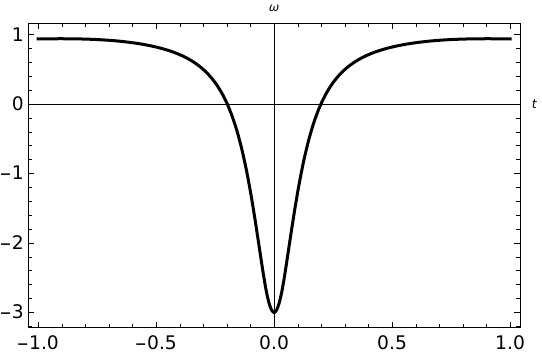}
\caption{Variation of EoS parameter with cosmic time for $\sigma=10 $,\; $\alpha=10$,\; $\beta_1=-0.5$,\; $Q_0=0.5$ in exponential modified gravity}
\end{figure}

 The EoS parameter in Fig. (18) shows that it is negative near the bounce, thereafter increases to positive values, leading to stiff matter.

\section{ Stability of de Sitter-like Solutions}

The field eq. (19) can be rewritten as
\begin{equation}
\label{e29}
3 H^2 =  \rho+ \rho_{mod}
\end{equation}

where $\rho_{mod}= - \frac{Q_0}{2} +54 \alpha_1 H^4 +18 \beta_1 \left(\dot{H}^2 +6 \dot{H} H^2 + 9 H^4\right) - 36 \beta_1 H \left( \ddot{H} + 6 H \dot{H} \right) $.
Assuming a universe field with isotropic fluid $p=\omega \rho$ in a modified gravity it yields
 \begin{equation}
\label{30}
3 H^2= -\frac{Q_0}{2}+54 \alpha_1 H^4 +18 \beta_1 \left(\dot{H}^2 +6 \dot{H} H^2 + 9 H^4\right) + 36 \beta_1 H \left( \ddot{H} + 6 H \dot{H} \right)   + \frac{\rho_o }{a^{\frac{1}{3(\omega +1)}}}
\end{equation}
where $\rho_o$ is an integration constant.
Now, ignoring the matter term, we look for a de Sitter solution which is given by
\begin{equation}
\label{e31}
H_o^2= \frac{1}{36 (\alpha_1 + 3 \beta_1)} \left( 1+ \sqrt{1+12 Q_0(\alpha_1+3 \beta_1)} \right)
\end{equation}
It has two solutions, for a positive Hubble parameter expanding universe and for a negative corresponds to a contracting universe.  The stability of the solutions 
will be studied. We consider a small perturbation in Hubble parameter as 
\begin{equation}
\label{e31a}
H(t)\rightarrow H_o+ \zeta (t) 
\end{equation}
without the matter part  and considering $\zeta <<H_o$:
we obtain
\begin{equation}
\label{e32}
\ddot{\zeta} + 3 H_o \dot{\zeta} - K_{\pm} \; \zeta =0
\end{equation}
where, $K_{\pm}= \frac{1}{36 \beta_1 H_o} \left[ 3 H_o^{2}  +\frac{Q_0}{2} -54\alpha_1 H_o^4 +162 \beta_1 H_o^4 \right]$
  Integrating eq. (\ref{e32}) we get
\begin{equation}
\label{e33}
 \zeta = C_1 e^{\lambda_1\;t }+ C_2   e^{\lambda_2\; t}
\end{equation}
 where the eigenvalues are $\lambda_1= K_{+}$ and  $\lambda_2= K_{-}$.
In the first case perturbation is growing exponentially and in the second case it is decaying leading  stable and unstable solutions respectively. A transition scheme  between stable and unstable particle contents in the universe 
may result a stable inflationary universe from bounce. This may be due to slowing down of the exponential expansion in the other branch. The quantum particles generated in the process yields a sufficient inflation.\\

\section{Discussion}

 In the paper, we explore bouncing universe in a modified gravity $f(Q,C)$ and also look for the role of the boundary term in the evolutionary behaviour of the universe.  Two different cosmological models namely,  power law model and exponential model are considered. The two models are probed in two different theories of gravity: linear in $Q$ and exponential in $Q$ are considered here for comparative study.
  We analyze the behaviors of the energy density and pressure for different values of the model parameters  $\alpha_1$, $\beta_1$, and $Q_0$ to obtain bounce universe.  In inflationary universe, it is not known when and how the universe entered into inflation. Before inflation, the universe was expanding slowly, which could not avoid singularity unless there exists a bounce phase. We note the following:
(i) The bounce universe with $q<0$ transits to $q>0$ at a later epoch during the expansion phase of the universe (Figs. (3), 4(b), (9) during power law expansion (PLE) but it is always negative and $q>-1$ for an emergent universe (Fig. (10)).   \\
(ii) The energy density is positive but the pressure is negative near the bounce as shown in Figs. 5(a), 5(b) with a different trend in Fig. 6(a) and 6(b) which is obtained In fig. 7(a) and 7(b)  the energy density increases and pressure decreases from a maximum value attained at the time bounce in exponential model in the absence of the quadratic boundary term but in the presence of the quadratic term  it is evident that near the bounce negative pressure increases to a region with  $p>0$ attains a maximum and thereafter decreases to $p=0$ as shown in Fig. 6(b) in the case of power law expansion,   and in Fig. 7(b) a symmetric bounce where energy raises from a minimum positive to attain a maximum and thereafter  decreases. SEC is violated \\
(iii) A negative  equation of state  in Fig. 4(a) obtained with linear $C$ but is found to differ with $C^2$ term where $\omega$ transits from negative value to a positive value in course of evolution (Fig. 8) during PLE, but we note that a positive $\omega$ right from bounce point decreases attains a  minimum with negative value and thereafter it remains negative (Fig.12). The Eos changes from $q< -1$ to $q=1$ as shown in Fig. (14).\\
(iv) In Fig (12) EoS parameter changes from $\omega >0$ to negative value for exponential expansion.\\
(v) For emergent universe model, NEC is always violated for exponential expansion as shown in Fig. 13(c) without quadratic boundary term, but for a power law expansion NEC is violated near the bounce thereafter it satisfies throughout as shown in Fig. 5(c), 6(c), 7(c) and 13(c). \\
(vi) The EoS parameter evolves from a negative value $\omega<0$ to a positive value $\omega>0$ away from   the bounce as shown in Fig. (14).\\
(vii) There is a single bounce with linear boundary term, but one more bounce is found to exist if $C^2$ term is added to the action.\\
(viii) The effect of quadratic boundary term in the gravitational action is considered for analyzing the validity of the energy conditions, namely,  null-energy condition, dominant-energy condition, and strong-energy conditions. Bouncing solution of the universe is probed with two different scale factors. We also study the stability of de Sitter-like solutions in the gravitational models, which is compatible with the expected results. The bounce cosmological model is stable and gives non non-singular universe at the origin.\\
(ix) In Fig 6(a) the energy density is positive and there will be a bounce near the t tends to zero, but we see that for different values of parameter $\alpha_1$  and $\beta_1=-0.5$ the peak of the graph decreases and there will be transition of energy density from positive to negative between $\alpha_1=2$ and $\alpha_1=3$. \\
(x) We also study the variation of energy density for different values of $\alpha_1$ and $\beta_1=0.0005$ in Fig 7(a), there will be positive energy density only for $\alpha_1=1$ and for other values of $\alpha_1$ it will be in the negative region.\\
(xi) It is evident that for the emergent universe model introducing quadratic term $c^2$, we get certain results for different values of coupling parameters $\alpha_1$ and $\beta_1$, in which NEC is obeyed and also violated (Figs 15(c) and 16(c)). In Fig 15(c), we get $\rho=0$ and $p=0$ at the bouncing point, which is equivalent to the vacuum case.\\
(xi) In our paper, we introduce two different forms of modified gravity: one is a linear form of modified gravity, and the other is the exponential form of modified gravity. In the first form of modified gravity, we introduce the power law model and emergent universe model, and in the latter case, we discuss the exponential form of modified gravity in terms of the power law model.\\
(xii) In the linear form of modified gravity for $\alpha=1$ and $\alpha_1=0$, $\beta_1=0$ in case of power law model Eos parameter($\omega$) and deceleration parameter ($q$) both are similar, and they show a single bounce near $t\rightarrow 0$, which shows a transition from declaration to acceleration of the universe and non singular nature of the universe (Figs 4(a) and (b)). But for non-zero values of coupling parameters, there will be a double bounce near $t\rightarrow 0$(Fig(8). In case of the emergent universe model for the linear form of modified gravity Eos- parameter($\omega$)shows a single bounce near $\omega>0$ but it eventually decreases( phantom region)(Fig.(14)). But in case of exponential form of gravity, introducing power law, we get an Eos parameter ($\omega$) which shows presence of stiff matter near the bounce.\\
(xiii) In the linear form of gravity, introducing power law we see that energy density is always positive for $\alpha_1=0$ and $\beta_1=0$  (Figs. 5(a) and (b))and non zero values of coupling parameters in Fig 6(a) but in case of exponential form of gravity we can see that energy density is positive  $\alpha<$ and $\beta_1<0$ , and $\alpha>0$ and $\beta_1<0$ (Fig 17(a)). Also, in linear model we get $\rho=0$ and $p=0$ near the bounce for emergent model shown in Figs. 15 (a) and (b) but for exponential form of gravity using power law model we get nonzero values of  $\rho$ and $p$ (Figs.(17(a) and (b)). NEC and SEC are violated near the bounce for both forms of modified gravity.
\vspace{1 cm}

{\bf Acknowledgment :}
BM is thankful to IUCAA Centre for Astronomy Research and Development (ICARD), NBU for extending research facilities for the project.
BCP would like to thank IUCAA for research facilities under Visiting Associateship Program and ANRF, Govt. of India for a research grant (F. No. CRG/2021/000183).


\begin{thebibliography}{}

 \bibitem{1}  
A. H. Guth, {\it Phys.Rev. } {\bf D 23}, 347 (1981); R.  Brout,  F.  Englert , E.  Gunzig,  A. A. Starobinsky, {\it Phys. Lett. } {\bf B 91}, 99 (1980); A. D. Linde, {\it Phys. Letts.}  {\bf      },  1983.
K. Sato,  {\it Mon. Not. Roy. Astron. Soc.} {\bf 195}, 467 (1981)

 \bibitem{2}   V. Mukhanov and G. Chibisov, {\it JETP Lett.} {\bf 33},  532  (1981)
 
  \bibitem{3}  R. H. Brandenberger,?Inflationary cosmology: Progress and problems,? {\it hep-ph/9910410}.  
  
  
  \bibitem{4}  N.  Arkani-Hamed,   H.  C.  Cheng,   M.  A.  Luty, S.  Mukohyama,   {\it JHEP} {\bf 0405}, 074  (2004);
  P. Creminelli, M. A. Luty, A. Nicolis and L. Senatore, {\it JHEP} {\bf 0612},  080  (2006)  

   \bibitem{BCP}  G. F. R. Ellis, J. Murugan and C. G, Tsagas, Class.
Quantum Grav. 21 (2004) 233; G. F. R. Ellis and R. Maartens, Class. Quantum Grav.
21 (2004) 223; J. Mulryne, R. Tavakol, J. E. Lidsey and G. F. R.
Ellis, Phys. Rev. D 71 (2005)123512.
 \bibitem{bcp}  S. Mukherjee, B. C. Paul, S. D. Maharaj and A.
Beesham, arXiv:gr-qc/0505103 (2005); B. C. Paul and A. Chanda, General Relativity and
Gravitation 51 (2019) 71; P. S. Debnath and B. C. Paul,
Mod. Phys. Letts. A 32 (2017) 1750216; B. C. Paul and
P. Thakur, Astrophys. \& Space Sci. 362 (2017) 73; S.
Ghose, P. Thakur and B. C. Paul, Mon. Not. Roy. Astro.
Soc. 421 (2012) 20; B. C. Paul, S. Ghose and P. Thakur,
Mon. Not. Roy. Astro. Soc. 413 (2011) 686; B. C. Paul
and A. Saha, Class. Quantum Grav. 27 (2010) 215004;
B. C. Paul, P. Thakur and S. Ghose, Mon. Not. Roy.
Astro. Soc. 407 (2010) 415.

\bibitem{bcp1}  B C Paul and A S Majumdar, Class. Quantum Grav. 32
(2015) 115001; B C Paul and A S Majumdar, Class. Quantum Grav. 35
(2018) 065001.
\bibitem{bcp2} B. C. Paul, {\it"Emergent universe in dimensions with dynamical wormholes" }{\it Eur. Phys. J.} {\bf  C 81}, 776 (2021). 

   
   \bibitem{5}   B. Elder, A. Joyce and J. Khoury,  {\it  Phys. Rev. } {\bf D 89}   044027 (2014); A.   Nicolis,   R.   Rattazzi   and   E.   Trincherini, {\it  Phys. Rev.} {\bf D 79}, 064036 (2009); P.  Creminelli,  K.  Hinterbichler,  J.  Khoury,  A.  Nicolis and E. Trincherini, {\it JHEP} {\bf 1302}, 006 (2013), {\it JHEP} {\bf 02} 006 (2013);      T.  Biswas,  T.  Koivisto  and  A.  Mazumdar,   {\it JCAP} {\bf 1011},  008  (2010).

  \bibitem{6}   S.  Dubovsky,   T.  Gregoire,   A.  Nicolis  and  R.  Rattazzi,  {\it  JHEP} {\bf 0603},  025  (2006).
  
  \bibitem{7}   V.    A.    Rubakov, {\it    Phys.   Usp. } {\bf 57},    128   (2014)
  
    \bibitem{8} S. Capozziello, V. De Falco, C. Ferrara, {\it Euro. Phys. J} {bf 83}, 915 (2023)
\bibitem {9}  S. Perlmutter, et al.,  {\it Measurements of $\Omega$ and $\Lambda$ from 42 High-Redshift Supernovae}  {\bf 517 (2)}, 565–586 (1999) 
\bibitem {10} A. G. Riess, et al.,{\it  Observational Evidence from Supernovae for an
Accelerating Universe and a Cosmological Constant, Astron. J.} {\bf 116 (3)}(1998) 
\bibitem {11} A. G. Riess, et al.,{\it  Type Ia supernova discoveries at z > 1 from the Hubble Space Telescope: Evidence for past deceleration and constraints on dark energy evolution, Astrophys. J.}{\bf  607 (2)} 665 (2004) 
\bibitem{12}  D. N. Spergel, {\it  et al.}, First Year Wilkinson Microwave Anisotropy Probe {\it Phys. J. Suppl. Series}{\bf 148 } 175 (2003). 
\bibitem {13}  T. Koivisto, D. F. Mota, {\it Dark energy anisotropic stress and large scale structure formation, Phys. Rev. } {\bf D 73 (8)} (2006). 
\bibitem {14}  S. F. Daniel, R. R. Caldwell, A. Cooray, A. Melchiorri, {\it Large scale structure as a probe of gravitational slip, Phys. Rev.} {\bf D 77 (10)} (May 2008). 
\bibitem {15} Y. Minami, E. Komatsu, {\it New Extraction of the Cosmic Birefringence from the Planck 2018 Polarization Data, Phys. Rev. Lett} {\bf 125 (22)} (Nov 2020). 
\bibitem {16} V. Sahni, A. Starobinsky, {\it THE CASE FOR A POSITIVE COSMOLOGICAL $\Lambda$TERM, Int. J. Mod. Phys.} {\bf D 09 (04)} 373–443 (2000).
\bibitem{17} S. M. Carroll, {\it The cosmological constant, Living Rev. Rel.} {\bf 4 (1)} (Feb 2001).
\bibitem{18} T. Padmanabhan, {\it Cosmological constant—the weight of the vacuum, Phys. Rept.} {\bf 380 (5-6)} 235-320(2003).
\bibitem{19} P. J. E. Peebles, B. Ratra, {\it The cosmological constant and dark energy Rev. Mod. Phys.} {\bf 75 (2)} 559-606 (2003).
\bibitem{20} T. Chiba, {\it Quintessence, the gravitational constant, and gravity, Phys.Rev.} {\bf D 60 (8)} (1999) 
\bibitem{21} L. Amendola, {\it Coupled quintessence, Phys. Rev.} {\bf D 62 (4)} (2000)
\bibitem{22} J. Martin, {\it Quintessence: a mini-review, Mod. Phys. Lett.} {\bf A 23 (17)}1252–1265(2008)
\bibitem{23} A. Kamenshchik, U. Moschella, V. Pasquier,{\it An alternative to quintessence, Phys. Lett.} {\bf B 511 (2-4)} 265–268 (2001).
\bibitem{24} M. C. Bento, O. Bertolami, A. A. Sen {\it Generalized chaplygin gas, accelerated expansion, and dark-energy matter unification, Phys. Rev.}{\bf D 66 (4)} (Aug 2002).
\bibitem{25} H. Benaoum, {\it Accelerated universe from modified chaplygin gas and tachyonic fluid, Universe} {\bf 8 (7)} 340 (2002). 
\bibitem{26} T. P. Sotiriou, V. Faraoni, {\it f(R) theories of gravity, Rev. Mod. Phys.} {\bf 82 (1)} 451–497 (2010). 
\bibitem{27} A. D. Felice, S. Tsujikawa, {\it f(R) Theories, Living Rev. Rel.} {\bf 13 (1)} (Jun 2010).
\bibitem{28} T. Harko, F. S. N. Lobo, S. Nojiri, S. D. Odintsov, {\it f(R, T) gravity, Phys. Rev.} {\bf D 84 (2)} (Jul 2011).
\bibitem{29} S. Nojiri, S. D. Odintsov, {\it Modified Gauss-Bonnet theory as gravitational alternative for dark energy, Phys. Lett.} {\bf B 631 (1)} 1-6 (2005).
\bibitem{30} G. Cognola, E. Elizalde, S. Nojiri, S. D. Odintsov, S. Zerbini, {\it Dark energy in modified Gauss-Bonnet gravity: Late-time acceleration and the hierarchy problem, Phys. Rev.} {\bf D 73} (2006).
\bibitem{31} B. Li, J. D. Barrow, D. F. Mota, {\it Cosmology of modified Gauss-Bonnet gravity, Phys. Rev.} {\bf D 76} (2007).
\bibitem{32} R. Maartens, K. Koyama, {\it Brane-world gravity, Living Rev. Rel.} {\bf 13 (1)} (Sep 2010). 
\bibitem{33} P. Brax, C. van de Bruck, A.-C. Davis,{\it Brane world cosmology, Rept. Prog. Phys.}{\bf 67 (12)} 2183–2231(2004). 
\bibitem{34} A. Wang, {\it Hoˇrava gravity at a lifshitz point: A progress report, Int. J. Mod. Phys.}{\bf D 26 (7)} (2017). 
\bibitem{35} A. Starobinsky, {\it A new type of isotropic cosmological models without singularity, Phys. Lett.}{\bf B 99} (1980).
\bibitem{36} S. Capozziello, V. F. Cardone,  V. Salzano, {\it Cosmography of f(R) gravity, Phys. Rev.} {\bf D 78 }(2008).
\bibitem{37} S. Capozziello, M. D. Laurentis, {\it The dark matter problem from f(R) gravity viewpoint, Ann. der Phys.} {\bf 524} 545–578(2012).
\bibitem{38} R. Myrzakulov, {\it FRW cosmology in F(R, T) gravity, Euro. Phys. J.} {\bf C 72 (11)} (Nov 2012).
\bibitem{39} P. Rudra, K. Giri, {\it Observational constraint in f(R, T) gravity from the cosmic chronometers and some standard distance measurement parameters, Nucl. Phys.} {\bf B 967} (2021).
\bibitem{40} B. C. Paul, A. Chanda, A. Beesham, S. D. Maharaj, {\it Late time cosmology in f(R, G)-gravity with interacting fluids, Class. Quantum Grav.}{\bf 39 (6)} (2022).
 \bibitem{41} S. Nojiri, S. D. Odintsov, O. G. Gorbunova, {\it Dark energy problem: from phantom theory to modified Gauss-Bonnet gravity, J. Phys.} {\bf A: Mathematical and General 39 (21)} 6627–6633 (2006).
\bibitem{42}E. D. Valentino, N. A. Nilsson, M.-I. Park, {\it A new test of dynamical dark energy models and cosmic tensions in hoˇr ava gravity, MNRAS} {\bf 519 (4)} 5043–5058 (2023).
\bibitem{43} T. Zhang, F.-W. Shu, Q.-W. Tang, D.-H. Du, {\it Constraints on hoˇrava–lifshitz gravity from GRB 170817a, Euro. Phys. J.} {\bf C 80 (11)} (Nov 2020). 
\bibitem{44} C. Pellegrini, J. Plebanski, {\it Tetrad fields and gravitational fields, Kgl.Danske Videnskab. Selskab, Mat. Fys. Skrifter}{\bf 2 (4)} (1963).
\bibitem{45} K. Hayashi, T. Shirafuji,{\it New general relativity, Phys. Rev.} {\bf D 19} 3524–3553 (1979).
\bibitem{46} E. V. Linder, {\it Einstein’s other gravity and the acceleration of the Universe}, Phys. Rev. D 81 (12) (Jun 2010).
\bibitem{47} J. W. Maluf, {\it The teleparallel equivalent of general relativity, Ann. der Phys.}{\bf 525 (5)} 339–357 (2013).
\bibitem{48} R. Aldrovandi, J. G. Pereira, {\it Teleparallel Gravity: An Introduction}, {\bf Vol. 173, Springer, New York}, 2013. 
\bibitem {49} S. Capozziello, V. F. Cardone, H. Farajollahi, A. Ravanpak, {\it Cosmography in f(T) gravity, Phys. Rev.} {\bf D 84 (4)} (Aug 2011).
\bibitem{50} R. Ferraro, F. Fiorini, {\it Modified teleparallel gravity: Inflation without an inflaton, Phys. Rev.}{\bf D 75} (2007) 084031.
\bibitem{51} R. Ferraro, F. Fiorini, {\it Born-Infeld gravity in Weitzenb¨ock spacetime, Phys. Rev.}{\bf D 78 (12)} (2008) 124019. 
\bibitem{52} G. R. Bengochea, R. Ferraro, {\it Dark torsion as the cosmic speed-up, Phys. Rev.} {\bf D 79 (12)} (2009) 124019. 
\bibitem{53} S. Capozziello, G. Lambiase, E. Saridakis, {\it Constraining f(T) teleparallel gravity by big bang nucleosynthesis: f(T) cosmology and BBN, Euro. Phys. J.} {\ C 77}  1–6 (2017).
\bibitem{54} A. Awad, W. E. Hanafy, G. Nashed, S. Odintsov, V. Oikonomou, {\it Constant-roll inflation in f(T) teleparallel gravity, JCAP} 2018 (07).
\bibitem{55} J. B. Jim´enez, L. Heisenberg, T. S. Koivisto, 
{\it Teleparallel palatini theories, JCAP} 2018 (08).
\bibitem{56} H. Chaudhary, U. Debnath, T. Roy, S. Maity, G. Mustafa, M. Arora, {\it Constraints on the parameters of modified Chaplygin-Jacobi and modified Chaplygin-Abel gases in f(T) gravity} (2024).
\bibitem{57} Y.-F. Cai, S. Capozziello, M. D. Laurentis, E. N. Saridakis, {\it f(T) teleparallel gravity and cosmology, Rept. Prog. Phys.}{\bf  79 (10)} (2016).
\bibitem{58} J. M. Nester, H.-J. Yo, {\it Symmetric teleparallel general relativity} (1999).
\bibitem{59} M. Adak, O. Sert, {\it A Solution to Symmetric Teleparallel Gravity} (2004).
\bibitem{60} M. Adak, M. Kalay, O. Sert, {\it LAGRANGE FORMULATION OF THE SYMMETRIC TELEPARALLEL GRAVITY, Int. J. Mod. Phys.} {\bf D 15 (05)} 619–634 (2006). 
\bibitem{61} M. Adak, O. Sert, M. Kalay, M. Sari, {\it SYMMETRIC TELEPARALLEL GRAVITY: SOME EXACT SOLUTIONS AND SPINOR COUPLINGS, Int. J. Mod. Phys.} {\bf A 28 (32)} (2013). 
\bibitem{62} J. B. Jim´enez, L. Heisenberg, T. Koivisto, {\it Coincident general relativity, Phys. Rev.} {\bf D 98 (4)} (Aug 2018). 
\bibitem{63} K. F. Dialektopoulos, T. S. Koivisto, S. Capozziello, {\it Noether symmetries in symmetric teleparallel cosmology, Eur. Phys. J.} {\bf C 79 (7)} (Jul
2019). 
\bibitem{64} J. B. Jim´enez, L. Heisenberg, T. Koivisto, S. Pekar, {\it Cosmology in f(Q) geometry, Phys. Rev.} {\bf D 101} (2020) 103507.
\bibitem{65} F. Bajardi, D. Vernieri, S. Capozziello, {\it Bouncing cosmology in f(Q) symmetric teleparallel gravity, Eur. Phys. J. Plus} {\bf 135 (11)} 1–14 (2020).
\bibitem{66} S. Mandal, D. Wang, P. K. Sahoo, {\it Cosmography in f(Q) gravity, Phys. Rev.} {\bf D 102} (2020) 124029. 
\bibitem{67} S. Mandal, P. Sahoo, J. Santos, {\it Energy conditions in f(Q) gravity,
Phys. Rev.} { \bf D 102 (2)} (2020) 024057. 
\bibitem{68} S. Arora, P. K. Sahoo, {\it Crossing Phantom Divide in f(Q) Gravity, Ann. der Phys.} {\bf 534 (8)} (jun 2022). 
\bibitem{69} J. Lu, X. Zhao, G. Chee, {\it Cosmology in symmetric teleparallel gravity and its dynamical system} (2019). 
\bibitem{70} R. Lazkoz, F. S. Lobo, M. Ortiz-Ba˜nos, V. Salzano, {\it Observational constraints of f(Q) gravity, Phys. Rev.} {\bf D 100 (10)}(2019) 104027.
\bibitem{71} F. K. Anagnostopoulos, S. Basilakos, E. N. Saridakis, {\it First evidence that non-metricity f(Q) gravity could challenge $\Lambda$CDM, Phys. Lett.}{ \bf B
822 }(2021) 136634. 
\bibitem{72} S. A. Narawade, B. Mishra, {\it Phantom Cosmological Model with Observational Constraints in f(Q) Gravity, Ann. der Phys.} {\bf 535 (5)} (apr 2023).
\bibitem{73} L. Atayde, N. Frusciante, {\it Can f(Q) gravity challenge $\Lambda$CDM?, Phys. Rev.} {\bf D 104 (6)} (Sep 2021). 
\bibitem{74} F. K. Anagnostopoulos, V. Gakis, E. N. Saridakis, S. Basilakos,{\it New models and big bang nucleosynthesis constraints in f(Q) gravity, Eur. Phys. J.}{\bf C 83 (1)} (2023) 58.
\bibitem{75} G. Gibbons and G. Horowitz,  {\it Sixth Marcel Grossrnann Meeting on General Relativity, Proceedings,
Kyoto, Japan, 1991} - edited by T. Nakamura (World Scientific, Singapore, 1992).

\bibitem{76} A.Samaddar and S.S.Singh {\it Matter bounce cosmological model in $f(Q,C)$ gravity theory, Mod. Phys. Lett. } {\bf A }(2025)
\bibitem{77} R. Bhagat, S. V. Lohakare,and B. Mishra {\it Exploring the Viability of f(Q, T) Gravity: Constraining Parameters with Cosmological
Observations. arXiv:2501.08666v1 [gr-qc]}  (2025)
 

\end{thebibliography}
\end{document}